\documentclass[aps,prl,superscriptaddress,amsmath,amssymb]{revtex4}
\usepackage{graphicx}% Include figure files
\usepackage{bm}% bold math
\begin{document}

%\preprint{APS/123-QED}

\title{$p$-Wave Cold Collisions in an Optical Lattice Clock}

\author{N. D. Lemke}\altaffiliation{also at Department of Physics, University of Colorado, Boulder, CO 80309, USA}
\affiliation{National Institute of Standards and Technology, Boulder, CO 80305, USA}
\author{J. von Stecher}
\affiliation{JILA, NIST and University of Colorado, Department of Physics, Boulder, CO  80309, USA}
\author{J. A. Sherman}
\affiliation{National Institute of Standards and Technology, Boulder, CO 80305, USA}
\author{A. M. Rey}
\affiliation{JILA, NIST and University of Colorado, Department of Physics, Boulder, CO  80309, USA}
\author{C. W. Oates}
\affiliation{National Institute of Standards and Technology, Boulder, CO 80305, USA}
\author{A. D. Ludlow}\email{Electronic address: ludlow@boulder.nist.gov}
\affiliation{National Institute of Standards and Technology, Boulder, CO 80305, USA}

\date{\today}
\begin{abstract}
%We study the frequency shift arising from cold collisions in lattice-confined fermionic Yb. Supported by a quantitative theoretical model, our measurements reveal that $p$-wave collisions are primarily responsible for the cold-collision shift.
We study ultracold collisions in fermionic ytterbium by precisely measuring the energy shifts they impart on the atom's internal clock states. Exploiting Fermi statistics, we uncover $p$-wave collisions, in both weakly and strongly interacting regimes. With the higher density afforded by two-dimensional lattice confinement, we demonstrate that strong interactions can lead to a novel suppression of this collision shift. In addition to reducing the systematic errors of lattice clocks, this work has application to quantum information and quantum simulation with alkaline-earth atoms.
\end{abstract}
\pacs{42.62.Eh; 34.50.-s; 06.30.Ft; 32.30.-r}
\maketitle

\nopagebreak

Ultracold alkaline-earth atoms trapped in an optical field are rich physical systems and attractive candidates for quantum information processing~\cite{Hayes2007,Reichenbach2007,Daley2008,Gorshkov2009}, quantum simulation of many-body Hamiltonians~\cite{Gorshkov2010,Cazalilla2009,Foss-Feig2010b,Foss-Feig2010,Hermele2009}, and quantum metrology~\cite{Katori2003,Ludlow2008,Lemke2009,LeTarget2006,Akatsuka2008}. In each case, interrogating many atoms simultaneously facilitates high measurement precision, but can also yield high atomic density and the potential for atom-atom collisions at lattice sites with multiple atoms. For quantum information and simulation, these interactions can be a key feature; for quantum metrology, %they are an incidental complication requiring careful effort to minimize. Either way, these interactions must be well understood.
however, they present an undesired complication. In either case, these interactions need to be well understood.

To limit interactions in lattice clocks, the use of ultracold, spin-polarized fermions was proposed to exploit the Fermi suppression of $s$-wave collisions while freezing out higher partial-wave contributions. However, small collision shifts have been measured in fermionic $^{87}$Sr~\cite{Ludlow2008,Campbell2009,Gibble2009,Rey2010,Yu2010,Swallows2011} and $^{171}$Yb~\cite{Lemke2009}. %In this work, we show that $p$-wave collisions are mainly responsible for the cold collision shift in the Yb clock.
%In this work,
Aided by the quantum statistics that govern the interactions of these fermionic atoms, we present a complete picture of the cold collisions in the Yb lattice clock by performing measurements with state-of-the-art precision together with a quantitative theoretical model. While with Sr it was found that $s$-wave collisions can occur in the presence of excitation inhomogeneity, %, arising from the thermal distribution of motional states in the lattice.
%In contrast, %Aided by the quantum statistics that govern the interactions of these fermionic atoms,
with Yb we highlight here the important role that $p$-wave collisions can play in lattice clock systems. Moreover, we demonstrate techniques for canceling the collision shift that could be used to vastly reduce the clock uncertainty.

\begin{figure}[htb]
\resizebox{12cm}{!}{
\includegraphics[angle=0]{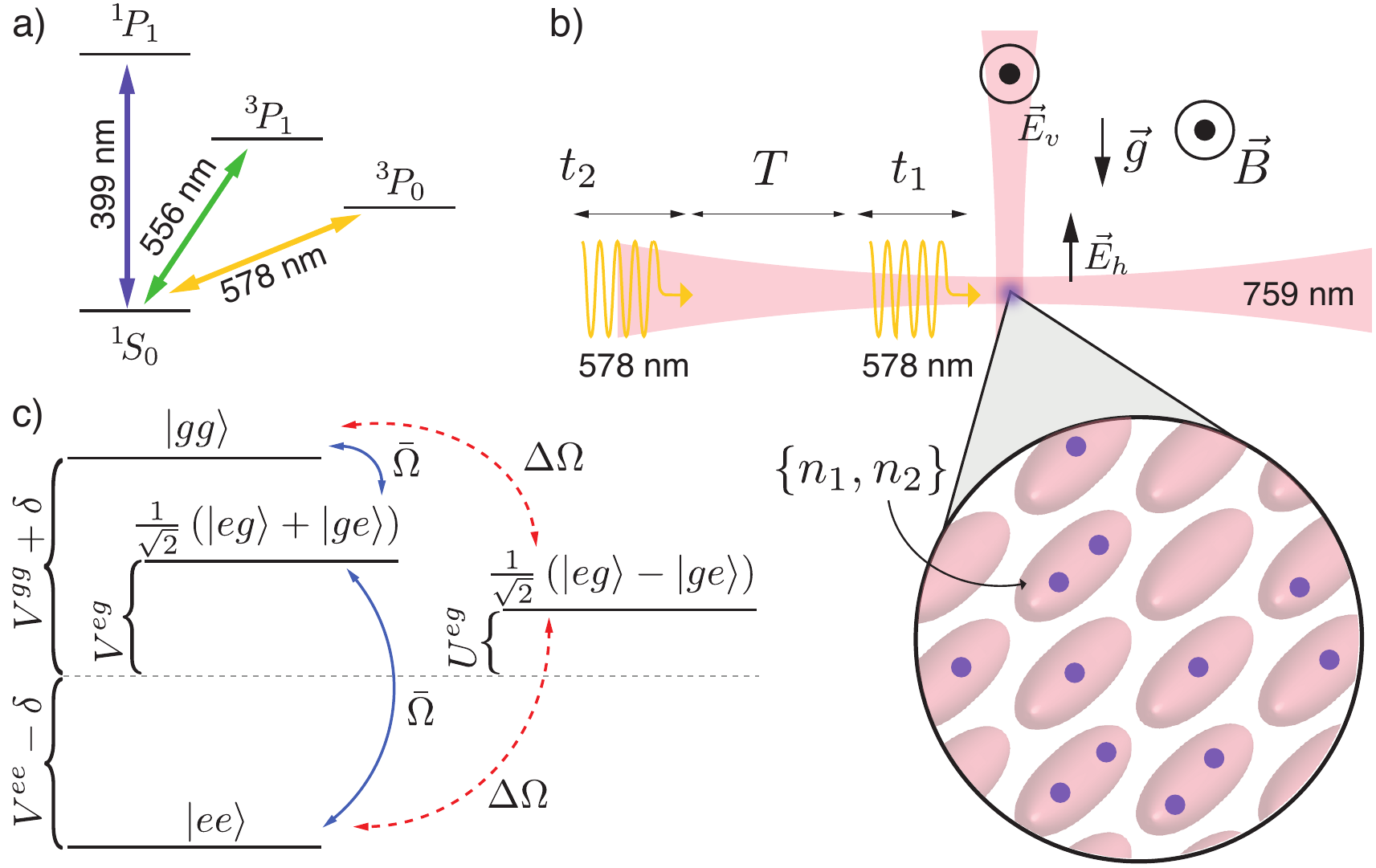}}
\caption{\label{Fig1} %\textbf{Experimental schematic and level diagrams.}
(a) Energy level diagram for Yb. (b) Schematic of the two lattices. Arrows indicate optical polarizations and magnetic field directions with respect to gravity. At the left, the Ramsey pulse sequence is entering: two pulses of times $t_{1,2}$ separated by dark time $T$ (not to scale). An inset shows a few 2-D lattice sites with 0 to 2 atoms per site; two atoms in one of the sites occupy axial motional states $n_1$ and $n_2$. (c) Energy level diagram for two atoms in the rotating frame: three triplet states and one singlet state, with interactions V and U, as in Eq.~(1).}
\end{figure}

%Because different interactions occur for different combinations of the atoms' internal states, precise knowledge of the relevant interaction is difficult.
To determine the nature of the collisions, here we use two-pulse Ramsey spectroscopy \cite{Ramsey1950} (Fig.~1). Provided the Ramsey pulse times $t_{1,2}$ are short compared to the dark time $T$, the vast majority of the collisions occur while the population is not being simultaneously driven by the laser field, simplifying the analysis of the interactions. The Ramsey technique also eliminates a strong laser-detuning-dependence on the excitation evolution and thus the interaction dynamics.  Finally, the Ramsey scheme offers the possibility to explore proposals for cancelation of the cold collision shift by tailoring the Ramsey pulses \cite{Gibble2009}.

In addition to measurements of the collision shift with atoms confined in a one-dimensional (1-D) optical lattice, we also show results from a two-dimensional (2-D) lattice, which offers several benefits.  First, with strong confinement in all but one dimension, the collisions can be treated with a 1-D model. Second, the higher number of lattice sites in a 2-D lattice reduces the lattice's overfilling (i.e., many atoms per site). Finally, the 2-D lattice offers stronger interactions at any lattice sites that are doubly occupied.%, allowing a different parameter space to be explored.

To model the collisions in a 2-D lattice, we begin by considering a simple case: two atoms in the same lattice site, populating axial vibrational modes $n_1$ and $n_2$ and the lowest transverse band \cite{Rey2010,Gorshkov2009,Gorshkov2010}. %We include both $s$- and $p$-wave interactions.
Assuming the vibrational quantum numbers are conserved during the collisions and laser interrogation, the Hamiltonian for the two-atom system can be written in a four-state basis set: %three pseudospin triplet configurations ($|gg\rangle$, $|ee\rangle$, $\left(|eg\rangle+|ge\rangle \right)/\sqrt{2}$) and one singlet ($\left(|eg\rangle-|ge\rangle \right)/\sqrt{2}$)
$|gg\rangle$, $|ee\rangle$, $\left(|eg\rangle+|ge\rangle \right)/\sqrt{2}$ (``triplet states'') and $\left(|eg\rangle-|ge\rangle \right)/\sqrt{2}$ (``singlet state'') \cite{Gibble2009,Swallows2011}. Here $g$ and $e$ are associated with the lowest ${}^1S_0$ and ${}^3 P_0$ electronic levels, respectively, which are coupled by the probe laser. The Hamiltonian in the rotating frame can be written in this basis as

\begin{equation}
H_T= \left(
              \begin{array}{cccc}
               \delta+ V^{gg}_{n_1 ,n_{2}}& 0 & \frac{\bar{\Omega}_{n_1,n_2}}{\sqrt{2}} & \frac{\Delta \Omega{n_1,n_2}}{\sqrt{2}} \\
               0 & -\delta + V^{ee}_{n_1 ,n_{2}}& \frac{\bar{\Omega}_{n_1,n_2}}{\sqrt{2}} & \frac{-\Delta \Omega_{n_1,n_2}}{\sqrt{2}} \\
                \frac{\bar{\Omega}_{n_1,n_2}}{\sqrt{2}}  & \frac{\bar{\Omega}_{n_1,n_2}}{\sqrt{2}} &  V^{eg}_{n_1 ,n_{2}} & 0 \\
                \frac{\Delta \Omega_{n_1,n_2}}{\sqrt{2}} &\frac{-\Delta \Omega_{n_1,n_2}}{\sqrt{2}} & 0 &   U^{eg}_{n_1 ,n_{2}}\\
              \end{array}
            \right).
\end{equation}

Here $\delta$ is the detuning  from the atomic transition, $\bar{\Omega}_{n_1,n_2}=(\Omega_{n_1}+\Omega_{n_2})/2$ is the average Rabi frequency for the two atoms, and $\Delta \Omega=(\Omega_{n_1}-\Omega_{n_2})/2$ is the difference in Rabi frequency. The dependence of the Rabi frequency on the axial vibrational
state is caused by any small projection of the probe beam along the axial direction. The terms $U^{eg}_{n_1 ,n_2}$ and $V^{\alpha\beta}_{n_1 ,n_2}$ give, respectively, the $s$- and $p$-wave interactions between an $\alpha=g,e$ and a $\beta=g,e$ atom \cite{SuppMat}. Because the atomic population is prepared in a single nuclear-spin state ($m_I= 1/2$ or $-1/2$), quantum statistics dictates that only the triplet states, which are invariant under particle exchange, are affected by $p$-wave interactions, while the singlet configuration interacts via $s$-wave only.

For short pulses and large Rabi frequencies, we can ignore interaction effects during the pulses. During the Ramsey dark time, the %effective
Hamiltonian describing the atom dynamics is diagonal in the singlet-triplet basis, and each state acquires just a phase. Consequently, after the second pulse is applied, we recover Ramsey fringes with a frequency shift. To compare with the 2-D lattice experiment, the shift must be integrated over the atomic distribution within an array of singly and doubly occupied lattice sites. Moreover, instead of choosing a specific set of vibrational modes $\{n_1,n_2\}$, we numerically calculate the appropriate thermal average over all possible modes \cite{SuppMat}. In the 1-D lattice, each  site is populated by many atoms, so the two-atom model is not directly applicable.  However, in the weakly interacting regime, a mean-field picture that approximates the many-atom interactions  by a sum of pairwise interactions provides a fair description, so we  use the two-atom Hamiltonian  with temperature-dependent, effective interaction parameters to model the multi-atom case. We compared this effective model with a numerically calculated N-body model and found qualitative agreement.

Our experimental procedures are similar to those described in \cite{Lemke2009}. After two stages of laser cooling (see Fig.~1(a)), atoms are trapped by the horizontal or vertical lattice for 1-D confinement, or by both lattices for 2-D operation. In the latter case, we filter away any atoms that are not fully confined at the intersection of the beams by adiabatically decreasing the power in one dimension to zero before turning it back up, then doing the same for the other lattice. Approximately $2.5\times 10^4$ atoms are trapped in the 1-D lattice, while $\tfrac{1}{5}$ as many remain in the 2-D lattice after filtering. In the 1-D lattice this corresponds to an estimated density of $\rho_1 = 3 \times 10^{11}/\text{cm}^3$ and $\sim$~20 atoms per site; for the 2-D lattice, we estimate that 25~$\%$ of the atoms are in doubly-occupied sites, for which the effective density is $\rho_2 = 4 \times 10^{12}/\text{cm}^3$, and fewer than 1~$\%$ of the atoms are in sites with more than two atoms. The trap frequencies in the lattice were typically 50-75~kHz in the strong direction(s) and 300-500 Hz in the weak direction(s). The lattice is tuned to the ``magic wavelength'' near $759~$nm, where the two clock states experience identical trapping potentials. We offset the frequencies of the two lattice beams by 2~MHz using acousto-optic modulators (AOMs), preventing any line-broadening from the vector Stark shift \cite{Porsev2004,Chin2001,Strabley2006,Lemke2009}.

With the atoms loaded in a lattice, we spin-polarize the sample by optical pumping to one of the spin states ($m_F=\pm 1/2$) with $556$~nm light; impurity in the spin-polarization is below 1~$\%$. %Next, all light sources except the lattice are switched off with mechanical shutters and AOMs, and a small magnetic field is applied to split the $\pi$ clock transitions by a few hundred hertz.
The clock light, pre-stabilized to a high-finesse optical cavity \cite{Jiang2011} to be resonant with the ${}^1S_0\rightarrow {}^3P_0$ clock transition, is switched on during the Ramsey pulses with an AOM. We interleave  high- and low-density clock conditions, each with its own set of integrators to lock the clock laser to the atomic transition and to average over both $m_F=\pm 1/2$ spin states. The collisional frequency shift is found by looking at the difference of the correction signals applied to the AOM divided by the difference in atom number, which is typically varied by changing the power to the $399$~nm slowing beam that opposes the atomic beam. These interleaved measurements have an instability of $\leq 1.5 \times 10^{-15}/\sqrt{\tau}$, for averaging time $\tau$ in seconds, allowing statistical error bars of $\sim 20$~mHz in just 2000~s. %When we report density shift numbers in Hz$/ \rho$, we average the atom number over the trap losses that occur during the spectroscopic time.

\begin{figure}[htb]
\resizebox{12cm}{!}{
\includegraphics[angle=0]{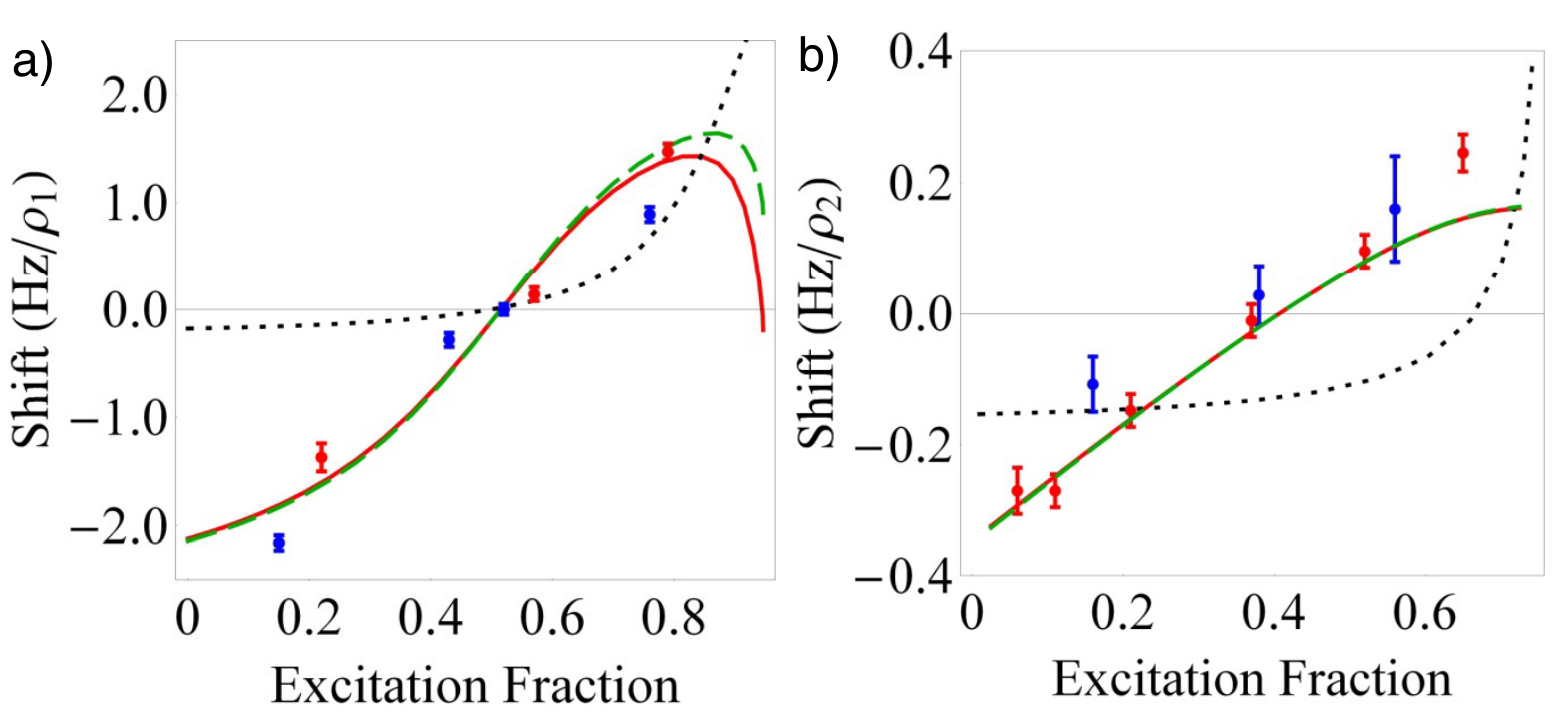}}
\caption{\label{Fig26} (a) Collision shift vs excitation fraction, 1-D lattice. Blue (red) points show experimental measurements in a vertical (horizontal) lattice with %density $\rho_1 \approx 3 \times 10^{11}/$cm$^3$,
temperature  $\mathcal{T}\sim 10$~$\mu$K and $\langle \Delta\Omega/\bar{\Omega}\rangle_\mathcal{T}=0.2$. Dashed black line gives an $s$-wave-only fit ($\langle U^{eg}\rangle_\mathcal{T}=-2 \pi \times 1.5$~Hz) from the mean-field model. Solid red line gives a $p$-wave-only fit with $\langle V^{eg}\rangle_\mathcal{T}=10 \langle V^{ee}\rangle_\mathcal{T}=-2 \pi \times 1.1 $~Hz. Long-dashed green line adds to this a small $s$-wave component ($\langle U^{eg}\rangle_\mathcal{T}=-2 \pi \times 0.6$). (b) Collision shift vs excitation fraction, 2-D lattice. Blue (red) points probe along the vertical (horizontal) lattice. % with density $\rho_2 \approx 4 \times 10^{12}/$cm$^3$ (for doubly occupied sites).
Dashed black line is an $s$-wave-only fit with $a_{eg}^-\approx-25$~$a_0$ ($a_0$ the Bohr radius); solid red line is a $p$-wave-only fit with $b_{eg}\approx-74$~$a_0$ and $b_{ee}^3=0.1 b_{eg}^3$. The long-dashed green line adds to this a small $s$-wave interaction $a_{eg}^-=-25$~$a_0$.}
\end{figure}

We first considered the collisional shift as a function of excitation fraction (i.e., the fraction in $|e\rangle$ during the Ramsey dark time).  The excitation fraction was varied by changing the Rabi frequency of the Ramsey pulses. The measured shift for atoms in a 1-D lattice is shown in Fig.~2(a) (blue and red points), and for atoms in a 2-D lattice in Fig.~2(b). For these data, the Ramsey pulse time is $t_1=t_2=1$~ms, and the dark time is $T=80$~ms.  For the 2-D lattice the black dashed and solid red curves give the numerically calculated shift using the $s$-wave scattering length ($a_{eg}^-$) and $p$-wave scattering volumes ($b_{eg}^3$ and $b_{ee}^3$) as fitting parameters. ($V^{gg}$ is taken to be zero, consistent with prior measurements \cite{Kitigawa2008}).  For the 1-D lattice, the curves are calculated from the mean-field approximation   with  the effective interaction parameters, which are  required  to be consistent with those used for the 2-D lattice calculations, varied for fitting.

Because the atoms are prepared in the triplet state $|gg\rangle$, $s$-wave interactions are allowed only in the presence of inhomogeneity ($\Delta \Omega_{n_1,n_2} \neq 0$)%\cite{Campbell2009}
, which transfers population to the singlet state. %\cite{Gibble2009,Rey2010}.
The collisions observed in $^{87}$Sr have been attributed to this type of interaction \cite{Campbell2009,Gibble2009,Rey2010,Yu2010,Swallows2011}. By contrast, $p$-wave interactions are fully allowed, provided there is sufficient collision energy to overcome the centrifugal barrier (expected to exceed 30~$\mu$K based on calculated van der Waals coefficients \cite{Dzuba2010}). %To distinguish the two, the black dashed (solid red) lines indicate calculations with purely $s$- ($p$-) wave interactions.
As shown in the figure, the $p$-wave interaction provides a much better description of the experimental data, as the shift induced by pure $s$-wave collisions is generally too small and does not exhibit the correct dependence on the excitation fraction. The shifts go through zero near an excitation fraction of 0.51 in the 1-D lattice and 0.4 in the 2-D lattice. Zero-crossings near 0.5  are readily understood if $V^{eg}$ dominates: by creating equal partial densities of ground and excited atoms, the energy shift on the two clock levels is the same, and the net shift is canceled. This effect could allow for vast reduction to the collisional shift in the Yb lattice clock. The deviation from a zero-crossing at exactly 0.5 in the 1-D case is consistent with a small $ee$ interaction ($b_{ee}^3= 0.1 b_{eg}^3$, with $b_{eg} \approx-74$~$a_0$ and $a_0$ the Bohr radius).

We investigated tunneling effects by measuring the shifts for both vertically and horizontally oriented 1-D lattices, exploiting gravity-induced suppression of the tunneling rate \cite{Lemonde2005}, but we observed no change in the data (Fig.~2(a)). We estimate the tunneling rate, thermally averaged over the lowest nine bands of the 1-D horizontal lattice, to be several hertz. In the vertical 1-D lattice and the 2-D lattice, this rate is suppressed by more than a factor of ten due to the energy offset between adjacent sites (arising from gravity \cite{Lemonde2005} and the Gaussian beam profile). However, the two highest bands of the lattice, which are populated with a few percent of the atoms, can have very high tunneling rates (several kilohertz) that are not suppressed by the energy offset. Because of this, the fraction of atoms that can tunnel during the spectroscopic time are 10~$\%$, 5~$\%$, and 10~$\%$ for the horizontal, vertical, and 2-D lattices, respectively. We do not see any appreciable difference between the shifts measured  in the horizontal and vertical lattices and thus conclude that tunneling does not play a significant role in the collision shifts.

We use   short Ramsey pulses ($t_{1,2} \sim 1$~ms) to avoid interaction effects during the pulse. But, for pulses shorter than the mean oscillation period in the trap, $\Omega \gtrsim \omega_i$, with $i$ the weakest trap direction, and laser-induced  mode-changing collisions are not necessarily suppressed. Nevertheless, we ruled out the relevance of  those processes by varying the pulse duration over a factor of ten  without  observing any substantial modification to the measured collision shifts \cite{SuppMat}. We looked for dependence of the collision shifts on the second pulse area \cite{Gibble2009}, but found no significant dependencies.

To further rule out $s$-wave interactions, we misaligned the probe beam to couple more strongly to the weak confinement axis of the lattice trap (Fig.~3(a)). Doing so introduces greater excitation inhomogeneity from the Ramsey pulse (in this case, up to a factor of 2.4) because the atoms are not tightly confined along this axis \cite{Campbell2009}. %The  excitation inhomogeneity $\Delta \Omega/\bar{\Omega}$  was increased by a factor of $2.4$, and
We expect the $s$-wave shift to depend quadratically on the inhomogeneity, yet the frequency shifts show no such dependence. This insensitivity is well explained by $p$-wave interactions, which depend only weakly on inhomogeneity at these levels (decreasing slightly as more population transfers to the singlet state). A small but non-zero $s$-wave interaction could balance this effect and may help explain the complete lack of dependence, as shown in the theory curves in Fig.~3(a).  The green long-dashed lines in Fig.~2(a,b) also show that adding a small but non-zero $s$-wave interaction is consistent with the observed collision shifts. Still, all of these considerations indicate that $p$-wave interactions play the dominant role in the cold collisions of $^{171}$Yb.

\begin{figure}[htb]
\resizebox{12cm}{!}{
\includegraphics[angle=0]{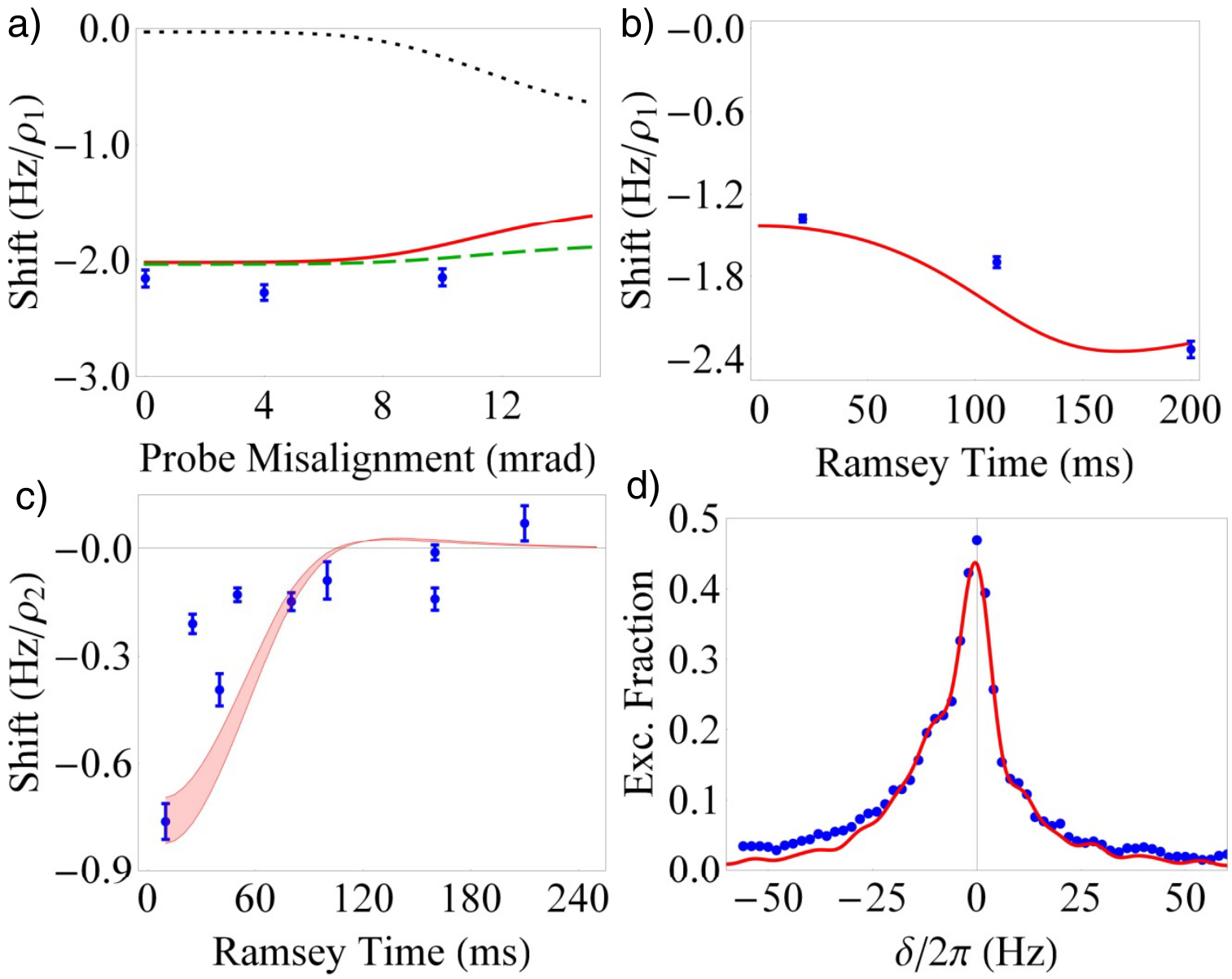}}
\caption{\label{Fig32}  (a) Collision shift vs probe misalignment angle (vertical 1-D lattice) for constant excitation fraction 0.12. Using the same parameters as Fig.~2(a), the dashed black line gives an $s$-wave-only fit, solid red line gives a $p$-wave-only fit, and the long-dashed green line has $s$- and $p$-wave terms. In the well-aligned case (0~mrad) there is a residual effective misalignment of $\sim5$~mrad due to the imperfect overlap between lattice and probe beams. (b) Collision shift vs Ramsey time, 1-D lattice, for excitation fraction $0.18$. Using the same parameters as Fig.~2(a), the solid-line gives a fit from the mean-field model.  (c) Collision shift vs Ramsey time, 2-D lattice, for excitation fraction $0.19\pm .03$. The shift crosses zero due to the periodic dependence of the shift on collisional phase, and is a signature of strong interactions. The model calculations (shaded region) use the same parameters as Fig.~2(b) for an excitation fraction range $0.19\pm .03$. (d) Asymmetric Rabi spectrum, 2-D lattice. The solid line is the prediction of the model, using the  number of doubly occupied sites as a fitting parameter.}
\end{figure}

Strong interactions emerge in the regime $V^{\alpha \beta}T \geq 1$.  A key observation revealing the operation of the 2-D lattice clock in the regime     $V^{ge}T \gg V^{ee}T \geq 1$  is the zero-crossing of the collision shift at a lower excitation fraction of 0.4, which deviates from the crossing at 0.5 predicted by the weakly interacting expression of the shift \cite{SuppMat}. The interaction strength also introduces additional dependencies on the Ramsey dark time. With weak interactions, the collision shift is independent of $T$, but with strong interactions the clock shift decays % as  ($\sim 1/T$)
with increasing $T$  due to the shift's sinusoidal  dependence on scattering phase \cite{SuppMat,Gibble2010}. We investigated this experimentally by varying the dark time $T$ and measuring collision shifts in the 1-D lattice (Fig.~3(b)), where the shift scales weakly with $T$, and the 2-D lattice (Fig.~3(c)), where the shift is strongly damped towards zero with increasing $T$. Yet a third signature of strong interactions is significant asymmetry in the clock transition spectrum. In Fig.~3(d) we show a Rabi spectrum ($t=120$~ms), taken under high density operation in the 2-D lattice, which shows an additional feature on the red side ($\delta<0$) of resonance. This asymmetry is density-dependent and barely observable in the 1-D lattice. In the 2-D lattice,  the interactions are sufficiently strong  ($V^{eg}t \geq 1 $) to introduce these asymmetric lineshape features beyond the transition linewidth. With yet higher density, it may be possible to spectrally resolve three features, one each for the $s$-wave-interacting singlet, the $p$-wave-interacting triplets,  and the non-interacting atoms in singly occupied lattice sites. Interaction-induced sidebands were recently reported in \cite{Bishof2011} and may be useful for quantum simulation applications.

In this Letter we have shown evidence for $p$-wave interactions in ultracold Yb confined in an optical lattice. Although lower atomic temperature yields reduced tunneling through the $p$-wave barrier, and thus a lower scattering cross-section, it also increases the atomic density of the confined atoms. For this reason, both $s$- and $p$-wave interactions may be potentially relevant for all optical lattice clock systems. Using the dependence of the measured shift on excitation fraction and Ramsey dark time, we have observed zero-crossings in the measured frequency shifts, which provide the metrological means to reduce the shifts to nearly negligible levels.

The authors gratefully acknowledge %optical frequency comb measurement support
assistance from Y. Jiang, S. Diddams, T. Fortier, and M. Kirchner and financial support from NIST, NSF-PFC, AFOSR, ARO, and DARPA-OLE.

%\bibliographystyle{apsrev}
%\bibliography{./pwaveBib_prl}

\newpage

\textbf{Supplementary Material}

\section{Two atom model}

Here we  consider two nuclear-spin-polarized fermionic atoms interacting via $s$-wave and  $p$-wave channels. The atoms are confined in  a tube with frozen transverse degrees of freedom (only the lowest vibrational transverse mode  is populated). Along the tube direction,  $\hat Z$, there is  a weak  harmonic confinement with frequency $\omega_Z$, and we will first  assume  the two atoms populate the  vibrational  modes $n_1$ and $n_2$.
In this case    there are only four states spanning the Hilbert space: the  triplet states   $|g g\rangle$, $|ee\rangle$, $(|e g\rangle + |ge\rangle)/\sqrt{2}$, and the singlet, $(|e g\rangle - |g e\rangle)/\sqrt{2}$. Here the convention used is that the left atom populates mode $n_1$ and the right atom mode $n_2$.

In the presence of a laser field with wave vector $\bm{k}=k_Y \hat Y + k_Z \hat Z$  and detuned from    the atom transition frequency by $\delta$, the two-atom Hamiltonian in the rotating frame  reduces to Eq. 1 in the main text. Here $\Omega_{n} = \Omega_0 L_{n}(\eta_Z^2) L_0(\eta_Y^2)  e^{-(\eta_Y^2 + \eta_Z^2)/2}$, with $\eta_i=k_i \sqrt{\frac{\hbar}{ 2m \omega_i}}$ the Lamb-Dicke parameter along the $i$ direction, $\Omega_0$ the bare Rabi frequency, and $L_n$ the Laguerre polynomial. $m$ is the atom mass.
 The $s$-wave and $p$-wave interaction parameters are $U^{eg}_{n_j ,n_{j'}}\equiv 4  \frac{\sqrt{m\omega_{X} \omega_{Y} \omega_{Z} }}{\sqrt{\hbar}} a_{eg}^- S_{n_j ,n_{j'}} $ and  $V^{\alpha\beta}_{n_j ,n_{j'}}\equiv 12 \frac{\sqrt{m^3\omega_{X} \omega_{Y} \omega_{Z}^3 }}{\sqrt{\hbar^3}} b_{\alpha,\beta}^3 P_{n_j ,n_{j'}}$. Here
$S_{n_j ,n_{j'}}$ and $P_{n_j ,n_{j'}}$  are geometric terms that take into account the spatial overlap between the atomic wavefunctions  of colliding atoms in modes ${n_j }$ and $n_{j'}$. They are given by $S_{n n' } =\frac{\int d\xi e^{-2\xi^2 }H^2_{n}(\xi)  H^2_{n'}(\xi) d\xi}{\sqrt{4^{n+n'} n!^2 n'!^2}}$ and  $P_{n n'} =\frac{\int d\xi e^{-2\xi^2 }\left(\frac{dH_{n}(\xi)}{d\xi}  H_{n'}(\xi)-H_{n}(\xi) \frac{ dH_{n'}(\xi)}{d\xi} \right)^2}{\sqrt{4^{n+n'} n!^2 n'!^2}}$, where  $H_n$ are Hermite polynomials. The dependence of $V^{\alpha\beta}$ and $U^{eg}$ on the vibrational mode encapsulates the temperature dependence of the interactions.

We ignore interactions during the pulses, and consequently the number of excited atoms after the first pulse is
\begin{eqnarray}
{N}_{n_1,n_2}^e(t_1)&=& 1-\cos(\bar{\Omega}_{n_1,n_2} t_1)\cos(\Delta{\Omega}_{n_1,n_2} t_1), \label{incoh}
\end{eqnarray}
During the dark time, the Hamiltonian is diagonal and each state acquires just a phase. After the second pulse, the excited state population is
\begin{eqnarray}
N_{n_1,n_2}^e(t_1,t_2)&=& A_{{n_1,n_2}}+ \mathcal{N}_{{n_1,n_2}} \cos[(\delta  -2\pi \Delta \nu_{{n_1,n_2}}^{ge})T ], \label{fringe}
\end{eqnarray}
with  $A_{{n_1,n_2}}$ an overall offset, $\mathcal{N}_{{n_1,n_2}}>0$ the fringe amplitude, and $\Delta \nu_{{n_1,n_2}}^{ge}$ the frequency shift. These quantities can be computed analytically.

In the 2-D lattice system an array of isolated tubes is populated with  mainly  one and two atoms per tube. If $N_0$ tubes are singly occupied and $N_1$ tubes doubly occupied,   then

\begin{eqnarray}
N^e_{n_1,n_2}(t_1,t_2)&=& \tilde A_{n_1,n_2}^e+ \cos[\delta T] \left(  \mathcal{ A}_{{n_1,n_2}}^{1}+\mathcal{ A}_{{n_1,n_2}}^{2e} \right) + \sin[\delta T] \mathcal{B}_{{n_1,n_2}}^e,
\end{eqnarray} with
{\tiny
\begin{eqnarray}
&&\tilde A_{n_1,n_2}^e=\sum _{i=1}^{N_1}[1-\sin \left(\Delta \theta _1^i\right) \sin \left(\Delta \theta _2^i\right) \sin \left(\theta _1^i\right) \sin \left(\theta _2^i\right) \cos \left(2 T
   \left(V_i^{\text{eg}}-U_i^{\text{eg}}\right)\right)-\cos \left(\Delta \theta _1^i\right) \cos \left(\Delta \theta _2^i\right) \cos \left(\theta _1^i\right) \cos
   \left(\theta _2^i\right)]\\&&\notag
   +\frac{1}{2}\sum _{i=1}^{N_0}[1-\sin \left(\Delta \theta _1^i\right) \sin \left(\Delta \theta _2^i\right) \sin \left(\theta _1^i\right) \sin \left(\theta _2^i\right) -\cos \left(\Delta \theta _1^i\right) \cos \left(\Delta \theta _2^i\right) \cos \left(\theta _1^i\right) \cos
   \left(\theta _2^i\right)];\\
&&\mathcal{B}_{{n_1,n_2}}^e =  \frac{1}{2}\sum _{i=1}^{N_1} \\&&\notag
-\cos \left(T C_i\right) \left[\cos \left(\Delta \theta _2^i\right) \sin \left(2 \theta _1^i\right) \sin \left(\theta _2^i\right) \sin \left(T B_i\right)+\sin \left(2
   \Delta \theta _1^i\right) \sin \left(\Delta \theta _2^i\right) \cos \left(\theta _2^i\right) \sin \left(T D_i\right)\right]\\&&\notag +2 \sin \left(\Delta \theta _1^i\right)
   \sin \left(\Delta \theta _2^i\right) \cos \left(\theta _1^i\right) \cos \left(\theta _2^i\right) \sin \left(T C_i\right) \cos \left(T D_i\right)+\cos \left(\Delta
   \theta _1^i\right) \cos \left(\Delta \theta _2^i\right) \sin \left(\theta _1^i\right) \sin \left(\theta _2^i\right) \left[\sin \left(T
   \left(V_i^{\text{ee}}-V_i^{\text{eg}}\right)\right)+\sin \left(T \left(V_i^{\text{eg}}-V_i^{\text{gg}}\right)\right)\right];\\
   && \mathcal{A}_{{n_1,n_2}} ^{1}=\frac{1}{2}\sum _{i=1}^{N_0}( \cos\theta_1^i \cos\theta_2^i \sin\Delta\theta_2^i \sin\Delta\theta_1^i +
 \cos\Delta \theta_1^i \cos \Delta \theta_2^i \sin\theta_2^i \sin\theta_1^i) \\
    &&\mathcal{A}_{{n_1,n_2}} ^{2e}=\frac{1}{2}\sum _{i=1}^{N_1}  \\&&\notag  \cos \left(\Delta \theta _2^i\right) \sin \left(\theta _1^i\right) \sin \left(\theta _2^i\right) \left[2 \cos \left(\theta _1^i\right) \sin \left(T B_i\right) \sin
   \left(T C_i\right)+\cos \left(\Delta \theta _1^i\right) \left(\cos \left(T \left(V_i^{\text{ee}}-V_i^{\text{eg}}\right)\right)+\cos \left(T
   \left(V_i^{\text{eg}}-V_i^{\text{gg}}\right)\right)\right)\right]+\\&&\notag\sin \left(\Delta \theta _2^i\right) \cos \left(\theta _2^i\right) \left[\sin \left(2 \Delta
   \theta _1^i\right) \sin \left(T C_i\right) \sin \left(T D_i\right)+\sin \left(\Delta \theta _1^i\right) \cos \left(\theta _1^i\right) \left(\cos \left(T
   \left(U_i^{\text{eg}}-V_i^{\text{ee}}\right)\right)+\cos \left(T \left(U_i^{\text{eg}}-V_i^{\text{gg}}\right)\right)\right)\right].
   \end{eqnarray}}  In the above equations, the dependence of the various parameters on the modes $\{n_1,n_2\}$ is omitted for simplicity but implied. The index $i$ runs over the set of doubly- and singly-occupied tubes. $(V_i^{\text{ee}}-V_i^{\text{gg}})=2 C_i$, $(V_i^{\text{ee}}\text{-2 }V_i^{\text{eg}}+V_i^{\text{gg}})=2B_i$, and $(-2
 U_i^{\text{eg}}+V_i^{\text{ee}}+V_i^{\text{gg}})=2D_i$. Due to the Gaussian profile of the laser beams, the trapping confinement varies from tube to tube. This  variation gives rise to tube-dependent  interaction parameters as well as tube-dependent Rabi frequencies  $\theta^i_s=t_s \bar{\Omega}^i$ and $\Delta \theta^i_s=t_s\Delta\Omega^i$. Here $s=1,2$.

If the atomic population is initially prepared in the excited state ($|e\rangle$) instead of the ground state ($|g\rangle$), then

\begin{eqnarray}
N^g_{n_1,n_2}(t_1,t_2)&=& \tilde A_{n_1,n_2}^g+ \cos[\delta T] \left(  \mathcal{ A}_{{n_1,n_2}}^{1}+\mathcal{ A}_{{n_1,n_2}}^{2g} \right) + \sin[\delta T] \mathcal{B}_{{n_1,n_2}}^g,
\end{eqnarray} with

{\tiny
\begin{eqnarray}
&&\tilde A_{n_1,n_2}^g=\sum _{i=1}^{N_1}[1+\sin \left(\Delta \theta _1^i\right) \sin \left(\Delta \theta _2^i\right) \sin \left(\theta _1^i\right) \sin \left(\theta _2^i\right) \cos \left(2 T
   \left(V_i^{\text{eg}}-U_i^{\text{eg}}\right)\right)+\cos \left(\Delta \theta _1^i\right) \cos \left(\Delta \theta _2^i\right) \cos \left(\theta _1^i\right) \cos
   \left(\theta _2^i\right)]\\&&\notag
   +\frac{1}{2}\sum _{i=1}^{N_0}[1+\sin \left(\Delta \theta _1^i\right) \sin \left(\Delta \theta _2^i\right) \sin \left(\theta _1^i\right) \sin \left(\theta _2^i\right) +\cos \left(\Delta \theta _1^i\right) \cos \left(\Delta \theta _2^i\right) \cos \left(\theta _1^i\right) \cos
   \left(\theta _2^i\right)];\\
&&\mathcal{B}_{{n_1,n_2}}^g =  \frac{1}{2}\sum _{i=1}^{N_1} \\&&\notag
\cos \left(\Delta \theta _2^i\right) \sin \left(\theta _2^i\right) \left[\sin \left(2 \theta _1^i\right) \sin \left(T B_i\right) \cos \left(T C_i\right)+\cos
   \left(\Delta \theta _1^i\right) \sin \left(\theta _1^i\right) \left(\sin \left(T \left(V_i^{\text{ee}}-V_i^{\text{eg}}\right)\right)+\sin \left(T
   \left(V_i^{\text{eg}}-V_i^{\text{gg}}\right)\right)\right)\right]+\\&&\notag\sin \left(\Delta \theta _2^i\right) \cos \left(\theta _2^i\right) \left[2 \sin \left(\Delta
   \theta _1^i\right) \cos \left(\theta _1^i\right) \sin \left(T C_i\right) \cos \left(T D_i\right)+\sin \left(2 \Delta \theta _1^i\right) \cos \left(T C_i\right)
   \sin \left(T D_i\right)\right];\\
   &&\mathcal{A}_{{n_1,n_2}} ^{2g}=\frac{1}{2}\sum _{i=1}^{N_1}  \\&&\notag
   \cos \left(\Delta \theta _2^i\right) \sin \left(\theta _1^i\right) \sin \left(\theta _2^i\right) \left[\cos \left(\Delta \theta _1^i\right) \left(\cos \left(T
   \left(V_i^{\text{ee}}-V_i^{\text{eg}}\right)\right)+\cos \left(T \left(V_i^{\text{eg}}-V_i^{\text{gg}}\right)\right)\right)-2 \cos \left(\theta _1^i\right) \sin
   \left(T B_i\right) \sin \left(T C_i\right)\right]\\&&\notag+\sin \left(\Delta \theta _2^i\right) \cos \left(\theta _2^i\right) \left[\sin \left(\Delta \theta _1^i\right)
   \cos \left(\theta _1^i\right) \left(\cos \left(T \left(U_i^{\text{eg}}-V_i^{\text{ee}}\right)\right)+\cos \left(T
   \left(U_i^{\text{eg}}-V_i^{\text{gg}}\right)\right)\right)-\sin \left(2 \Delta \theta _1^i\right) \sin \left(T C_i\right) \sin \left(T D_i\right)\right].
\end{eqnarray}}

In both cases the shift is given by \begin{eqnarray}\langle\Delta \nu \rangle_{\mathcal {T}}=\frac{ \arctan\left (\frac{\langle \mathcal{B}_{{n_1,n_2}} \rangle_\mathcal {T}}{\langle\mathcal{A}_{{n_1,n_2}} ^{2}+\mathcal{A}_{{n_1,n_2}} ^{1}\rangle_\mathcal{T}}\right)+ \pi p} {2\pi T}\label{shi},\end{eqnarray} with $ \langle \rangle_\mathcal{T}$ denoting a thermal average. $p$ is an integer that needs to be chosen so that  during the $g \to e$ ($e\to g$) interrogation  the total number of atoms driven to $e$ ($g$)  reaches a maximum value at  $\delta=2\pi \langle\Delta \nu \rangle_{\mathcal {T}}$, instead of a minimum. It also ensures that the shift  is a smooth function of $T$. If $p$ is not correctly chosen the shift  can jump discontinuously instead of becoming smoothly  displaced outside the first Ramsey fringe.  For the experimental regimes  described here, no discontinuity occurs and we can set $p=0$.

An important point to emphasize is the different dependence of the shift on $T$ in the weakly and strongly interacting regimes. In the weakly interacting  regime, $T V^{\alpha,\beta} \ll 1$ and $T U^{\text{eg}}\ll 1$,  the  $\mathcal{B}$ term provides the leading contribution, because it exhibits a linear dependence on interactions $\mathcal{B}\propto (T V^{\alpha,\beta} , T U^{\text{eg}})$. This results in a $T$-independent collision shift. On the other hand, in the strongly interacting regime, $(T V^{\alpha,\beta},T U^{\text{eg}}) \gg 1$, both terms  $\mathcal{B}$ and   $\mathcal{A}$ exhibit a nontrivial sinusoidal dependence on the various interaction parameters. This implies that   as $T$ increases,  both  $\mathcal{B}$ and   $\mathcal{A}$  exhibit faster but bounded oscillations and therefore, according to Eq. \ref{shi},  on average the shift decays  as $1/T$.

\section{ Interaction sidebands: Rabi Spectroscopy}

We consider now the case
%Consider now the situation
in which the system is interrogated during a long single Rabi pulse of duration $t$. In this situation, both interaction- and laser-driven terms must be accounted for at the same time during the dynamical evolution. For $s$-wave-dominated collisions, reaching the strongly interacting regime $U^{\alpha\beta} t > 1$ can lead  to a suppression of the clock shift \cite{Rey2010,Swallows2011,Bishof2011}. This is not necessarily the case for $p$-wave-dominated collisions, in which atoms interact even in the initially populated triplet manifold.

Moreover, if $(V^{eg}-V^{gg}) t > 1 $ and the atoms start in the triplet $|gg\rangle$ state, there is a large gap that they must overcome if $\delta t\ll 1$ to populate the  $\frac{1}{\sqrt{2}}(|ge\rangle+|eg\rangle)$ state. This means that only when $\delta= V^{eg}-V^{gg}$ do the two states become resonant, and  population  will be transferred between  $|gg\rangle$ and  $\frac{1}{\sqrt{2}}(|ge\rangle+|eg\rangle)$. Since   $|gg\rangle$ is coupled to  $|ee\rangle$ by  a second-order process,  population of  $|ee\rangle$ is  energetically suppressed. Only if   $\delta= \frac{C_{n_1,n_2}^i}{2}$  is there a two-photon resonance, but even in this case the overall amplitude is small since it is  proportional to $\bar{\Omega} ^2/ B_{n_1,n_2}$.

From these considerations, one can  write the lineshape of  the 2D lattice  array as
\begin{eqnarray}
N^{e}(T,\delta)&=& \left\langle \sum_{i=1}^{N_1} \rho[\sqrt{2} \bar{\Omega}_{n_1,n_2}^i, \delta -V^{eg,i}_{n_1,n_2} +V^{gg,i}_{n_1,n_2} ]
+\sum_{i=1}^{N_1}\rho\left [\sqrt{2}  \frac{(\bar{\Omega}_{n_1,n_2}^i)^2}{B_{n_1,n_2}^i} ,\delta  -\frac{C_{n_1,n_2}^i}{2}\right]\notag \right.\\&&\left.+ \sum_{i=1}^{N_1}\rho[\sqrt{2}\Delta{\Omega}_{n_1,n_2}^i,\delta -U^{eg,i}_{n_1,n_2}+V^{gg,i}_{n_1,n_2}]+\sum_{i=1}^{N_0}\rho[{\Omega}_{n_1}^i,\delta] \right \rangle_\mathcal{T}\end{eqnarray}with $\rho[x,y]=\frac{x^2}{x^2+y^2} \sin^2[\sqrt{x^2+y^2}t /2 ]$.
For comparison with experiment, we need to average over the tube array and   evaluate  thermal averages. If it were possible to resolve the various peaks we would expect the line shape to have three  single-photon resonances: One at $\delta=0$ coming from single occupied tubes, one $p$-wave related at $\delta \propto \langle V^{eg}-V^{gg} \rangle_{\mathcal{T}}$,  and one at $\delta \propto \langle U^{eg} -V^{gg}\rangle_{\mathcal{T}}$ due to $s$-wave collisions.

To obtain the excitation fraction and the interaction sidebands, we need to properly integrate over the spatial degrees of freedom of the lattice.
The characteristic harmonic oscillator frequencies $\omega_i$ ($i=X,Y,Z$) of each lattice site vary smoothly with the position of the lattice site modifying the Lamb-Dicke and interaction parameters. The variation of the trapping frequencies is obtained from the lattice potential
\begin{equation}
V(X,Y,Z)=-V_{X,0} \cos^2\left(\frac{\pi  X}{a}\right) e^{-\frac{2 \left(Y^2+Z^2\right)}{W^2}}-V_{Y,0} \cos^2\left(\frac{\pi Y}{a}\right) e^{-\frac{2 \left(X^2+Z^2\right)}{W ^2}},
   \end{equation}
where $a$ is the lattice spacing, $W\sim$35~$\mu$m is the beam waist, and $V_{X,0}$ and $V_{Y,0}$ are the lattice depths.  We extract the characteristic harmonic oscillator frequencies of a given site by expanding the lattice potential at the center of it. Assuming a uniform occupation of the sites in the center of the lattice, we transform the spatial integration into an integration over the relevant range of trapping frequencies weighted by the number of sites with such trapping frequencies.  For each set of trapping frequencies, a thermal average is carried out over both the harmonic oscillator modes and lattice vibrational modes.

This procedure is significantly simplified by noting that the main correction of the spatial average comes from the changes in the interaction terms. Thus,  changes in the Rabi frequencies are  accounted for only by using an average value, allowing  us to   transform the integration over trapping frequencies into an integration over interaction strengths.

\section{Other considerations}

We have restricted so far our analysis  to an effective one-dimensional system with at most two atoms per tube. For more than two atoms per tube the spin model is no longer diagonal in the collective spin basis, since the spin Hamiltonian is not SU(2) symmetric. This implies that for more than two atoms and outside the weakly interacting regime  our calculations are not exact. However,  we have performed numerical calculations for up to five  atoms per tube that show that the two-atom model can qualitatively predict the many-atom case, but with renormalized interaction parameters.

In the 1-D lattice there are two weakly confined directions and  multiple atoms per lattice site. Under these conditions, interaction-induced mode-changing collisions, which are not accounted for in our two-atom model \cite{Rey2010}, have to be included. We have studied the role of those processes by numerical evaluation of a more general multi-mode Hamiltonian and found that it mainly leads to a renormalization of the model parameters. All of these considerations  justify the validity   of the two-atom model  for the description of the 1-D lattice.

\section{Experimental data on various pulse durations}
\begin{figure}[htb]
\resizebox{12cm}{!}{
\includegraphics[angle=0]{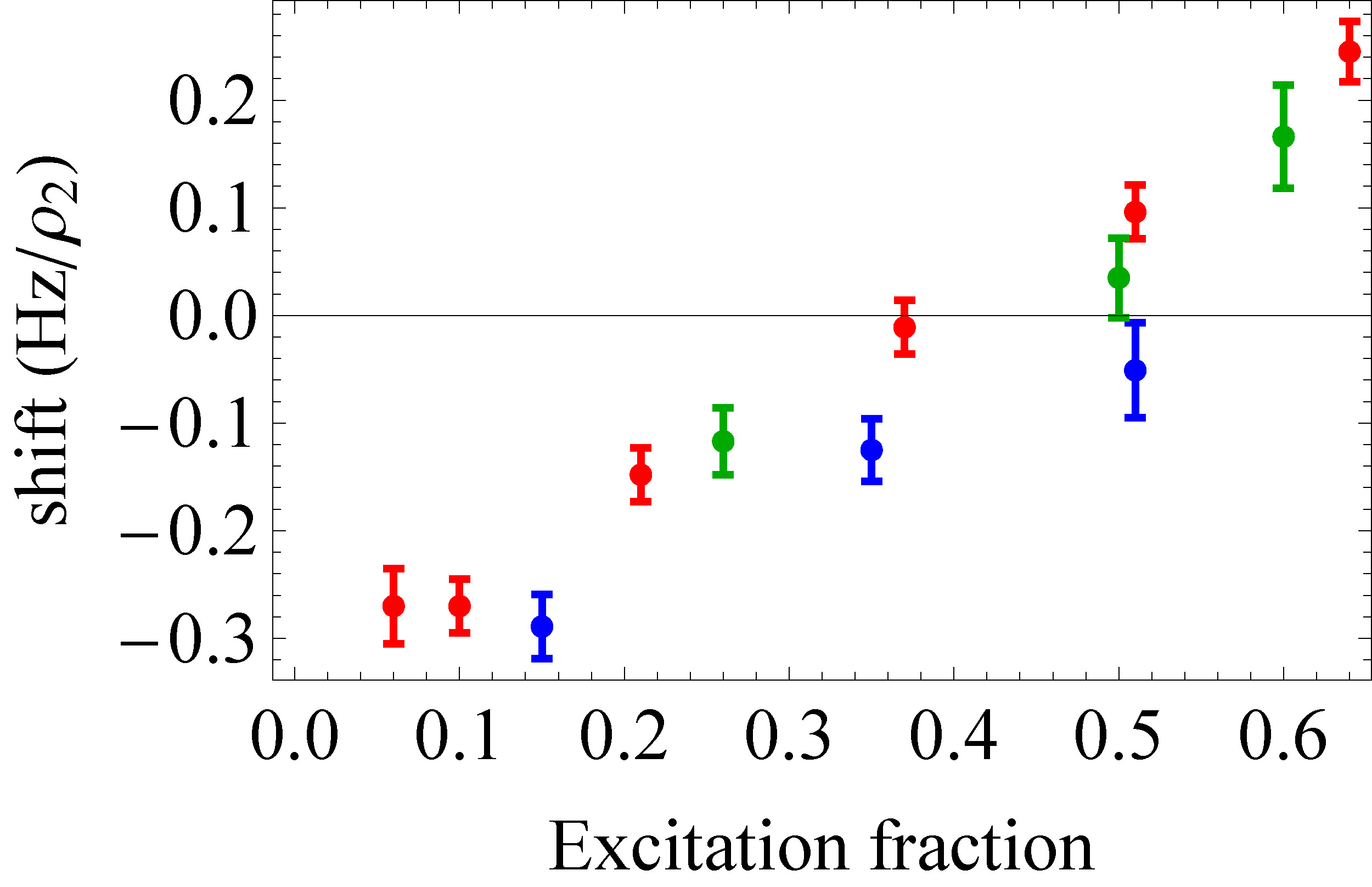}}
\caption{\label{Fig4} We study laser-driven mode-changing collisions by measuring the collisional frequency shift vs. excitation fraction for various Ramsey pulse times in the 2-D lattice. The blue, red, and green points correspond to $t_{1,2} = \{0.5, 1, 5 \}$~ms, respectively. Because we observe no substantial trend, we conclude that these processes do not play an important role.}
\end{figure}

We use   short Ramsey pulses ($t_{1,2} \sim 1$ms) to avoid interaction effects during the pulse. However, for pulses shorter  than the  mean oscillation period in the trap, $\Omega \gtrsim \omega_i$ with $i$ the weakest trap direction, and laser-induced  mode-changing collisions (not included in the two-atom model) are not necessarily suppressed. To explore the role of laser-induced mode-changing collisions, we measured the collision shift as function of excitation fraction in the 2-D lattice for various pulse durations. As shown in Figure~\ref{Fig4}, we varied the pulse time $t_{1,2}$ by one order of magnitude but found no substantial change in the measured shifts. Thus we conclude that these processes do not play an important role.

\end{document}